\documentclass{appolb}
\usepackage{graphicx}
\usepackage{setspace}

\begin{document}

\title{Multiscaling edge effects in an agent-based money emergence model}

\author{Pawe\l{} O\'swi\c ecimka$^{a}$, Stanis\l{}aw Dro\.zd\.z$^{a}$, Robert G\c ebarowski$^{b}$, Andrzej Z.
G\'orski$^{a}$, Jaros\l{}aw Kwapie\'n$^{a}$
\address{$^a$ Complex Systems Theory Department, Institute of Nuclear Physics, Polish Academy of Sciences, ul.
Radzikowskiego 152, 31-342 Krak\'ow, Poland\\
$^b$ Cracow University of Technology, Faculty of Physics, Mathematics and Computer Science, Warszawska 24, 31-155
Krak\'ow, Poland}}

\maketitle

\begin{abstract}
An agent-based computational economical toy model for the emergence of money from the initial barter trading,
inspired by Menger's postulate that money can spontaneously emerge in a commodity exchange economy, is extensively
studied. The model considered, while manageable, is significantly complex, however. It is already able to
reveal phenomena that can be interpreted as emergence and collapse of money as well as the related competition effects.
In particular, it is shown that - as an extra emerging effect - the money lifetimes near the critical threshold value
develop multiscaling, which allow one to set parallels to critical phenomena and, thus, to the real
financial markets.
\end{abstract}


\section{Introduction}
Extreme complexity of a phenomenon commonly termed as {\em money} stems from several factors.
Viewed in contemporary terms as a foreign exchange (Forex) market it can be considered the world's
largest and most important financial market, entirely decentralized, crossing all the countries,
with the highest daily trading volume extending to trillions of US dollars.
There is plenty of evidence \cite{kwapien2012} that the Forex's dynamics is more complex
than that of any other market's. The absence of an independent reference frame makes the absolute currency pricing
virtually impossible and a given currency's value is expressed by means of some other currency, which, in turn, is
also denominated only in currencies. Moreover, in global terms the Forex market is exposed
to current situation on other markets in all parts of the world, which makes it particularly sensitive and
unpredictable. These facts together with other Forex specific relationships like the triangle rule
\cite{aiba2002,drozdz2010} that links mutual exchange rates of three currencies are among the factors
responsible for a highly convoluted structure \cite{drozdz2007,kwapien2009,yakovenko2009} of Forex.

Complexity of money as a general medium of exchange that allows to avoid difficulties of barter trade requiring
a "double coincidence of needs" and constitutes a measure of and a store of value, roots back even to its beginnings.
Menger \cite{menger1892} proposed that money can spontaneously emerge in a commodity exchange economy. Accordingly, each
commodity is characterized by its own marketability reflecting its status in the market. Money is a commodity, which
through the process analogous to the physical spontaneous symmetry breaking \cite{bak2001} receives
a very high marketability and thus a special status of a medium of exchange \cite{kiyotaki1985}.
Katsuhito \cite{katsuhito1993,katsuhito1996} developed a model along the original  Menger's idea and demonstrated
that money is governed by a kind of the so-called bootstrap mechanism, i.e., it is accepted at any place at any time
because it is in a position of money. Agent based variants of such a model have further been studied by Yasutomi
\cite{yasutomi1995,yasutomi2003} and G\'orski et al. \cite{gorski2010} who by numerical simulations demonstrated that
they are able to reveal money emergence as well as its collapse. In general, agent based
models \cite{samanidou2007,chakraborti2011} find an increasing number of promising applications in various areas of
economics \cite{yasutomi1995,lux1999,iori2002,krawiecki2002,yang2008,denys2013} and in social sciences
\cite{kacperski1996,helbing2000,sznajd2000,grabowski2006,gawronski2011} and the need for this kind of approaching the
related phenomena, especially in an economic context, is being expressed more and more forcefully
\cite{bouchaud2008,farmar2009,lux2009}. For all these reasons in the present contribution, we further pursue simulations
based on an agent based model which from an initial barter trading is able to spontaneously elevate one of the
commodities to the money status. We in particular focus on the transition region between the homogenous commodities and
emerging money phases with the aim to identify the complexity characteristics of this transition in terms of the
multifractal scaling. Below we list the main ingredients of the model used.

\section{Model}
In the model \cite{yasutomi1995,gorski2010} we have $N$ agents, each agent producing one type of good
enumerated by $k = 1, \ldots, N$. For the sake of simplicity, we assume that the agent number $k$ is producing a good
type denoted by $k$. The elementary interaction of two agents ("transaction") consist of several
steps including search of the co-trader, exchange of particular goods, change of the agent's
buying preferences and finally the production and consumption phase. A sequence of $N$ consecutive transactions is
called a {\em turn}. In a single turn each of the $N$ agents has chance to take part in exchange of goods, production,
and consumption.
To each agent, say $k$, there are attributed three $N$-dimensional vectors. The possession vector, $P^{(k)}_i$, $i = 1,
\ldots, N$, with non-negative integer components that denote how many units of the $i$-th good has the $k$th agent
at the moment. The demand vector, $D^{(k)}_i$, is actually a "shopping list", i.e. it counts how many goods of the type
$i$ the
agent $k$ is going to buy. Finally, the "world view" vector, $V^{(k)}_i$, with non-negative real
components is related to the $k$th agent's shopping preferences.
These preferences are evolving with time, depending on the preferences of the other agents
(co-traders) as well as according to success of the previous transaction of the trader.
The vector $V^{(k)}_i$ is normalized according to
\begin{equation}
\label{Vnormalization}
\sum_{i=1}^{N} V^{(k)}_i = N, \quad \forall k,
\quad 0 \le V^{(k)}_i \le N
\ .
\end{equation}
The higher is the value of $V^{(k)}_i$, the more willing to buy the good $i$ is the  agent $k$. In addition, in each iteration 
for any $k$th agent there is a randomly attributed integer $w^{(k)}$ equal to the number $j = 1, \ldots, N$ pointing
a good produced by the $j$th agent that the $k$th agent urgently needs at the moment. 
Such a good is included in the shopping list independently of $k$th agent preferences ($V^{(k)}_i$).
The other goods at the shopping list will be added depending on the values of the "world view" (preference) vector.
In particular, if the value of a component $i$ of the vector $V_i$ is greater than the only external parameter of the
model, $Thresh \in [0, N]$, the good $i$ will be added to the shopping list.

The model algorithm is defined by the following steps:

\noindent{\bf Step 1.} An agent ("trader") $k$ is chosen randomly.

\noindent{\bf Step 2.} The trader $k$ chooses a co-trader (say, agent $l$)
who has the largest amount of wanted good, $w^{(k)}$.

\noindent{\bf Step 3.} Both traders check what they have and what they want.

\noindent{\bf Step 4.} The traders exchange their views.
At first, they increase the value of component $V^{(n)}$ ($n=l,k$) by $1.0$
if their previous demands were not satisfied, i.e.

\noindent $D^{(n)}_j > 0 \Longrightarrow  V^{(n)}_j \to V^{(n)}_j + 1$
($n=l,k$).

\noindent Then both traders accept an averaged view:

\noindent $V^{(n)}_j \to [V^{(k)}_j + V^{(l)}_j]/2$ ($n=l,k$).

\noindent Finally, the new views are re-normalized
according to the condition (\ref{Vnormalization}).

\noindent{\bf Step 5.} The traders create their "shopping list"
i.e. they decide what they want to buy. For the trader $k$:

\noindent if  $P^{(l)}_j > 0 \wedge (w^{(k)} = j \vee V^{(k)}_j > Thresh)$
$\Longrightarrow$ $D^{(k)}_j = P^{(l)}_j$,  otherwise   $D^{(k)}_j = 0$.

\noindent The same is done symmetrically ($k\leftrightarrow l$) by
the co-trader $l$ and for all types of goods, $j = 1,\ldots, N$.

\noindent{\bf Step 6.}  The exchange procedure. The traders "buy" (exchange)
goods according to their shopping lists $D^{(n)}_j$, $n=l,k$, $j = 1, \ldots,
N$.
If total amount of goods on both their shopping lists (demands) is identical,
$\sum_j D^{(k)}_j = \sum_j D^{(l)}_j$,
then their demands are fully satisfied and the shopping lists are zeroed.

\noindent If the shopping list of one trader (say, $k$) is bigger then the
shopping list of his co-trader, all demands of trader $k$
cannot be satisfied. Hence, after the exchange the vector $D^{(k)}_j$
will have non-zero components for the goods that could not be bought.
In this case the trader with a larger shopping list ($k$) can satisfy his
demands partially only. In particular, he selects from his co-trader one unit of good $j$ with the
smallest component $D^{(k)}_j$ (i.e. the agent prefers to get more rare goods).
This procedure is repeated unit by unit until the shopping list $D^{(l)}_j$ is zeroed.

\noindent If one of the traders has empty shopping list (all components are zero), there is no exchange at all and the
whole transaction is finished without any exchange. Notice, however, that in spite of this, the update of the
world view vectors was already done (Step 5). Also during this step, the possession vectors $P^{(k)}, P^{(l)}$
of both traders are updated.

\noindent{\bf Step 7.} The final step consists of consumption and production.
The traders $k, l$ consume goods specified by the variable $w^{(k)}, w^{(l)}$,
respectively. Then, if $P^{(k,l)}=0$, the traders produce one unit of the good
${k, l}$. Finally, we choose new wants for the traders: new values for the variables $w^{(k)}, w^{(l)}$ are randomly
selected $w^{(k)}\neq k$.
This ends the elementary transaction process.

The initial conditions are the following: $P^{(k)}_j = \delta_{kj}$,
$D^{(k)}_j = 0$, $V^{(k)}_j = 1$, and $w^{(k)}$ are chosen randomly
from the set $\{ 1, 2, \ldots, N\}$ for each $k$.
In particular, the initial shopping list is empty and the "views"
of all traders are identical and equally distributed for all goods.

In the model money is defined as the good satisfying the following
four conditions:
(i) that is most wanted by all agents,
i.e. $\sum_k V^{(k)}_j/N$ is maximized ($=V_{max}$) for the value of $j$
that corresponds to the good that plays the role of money;
(ii) the total trade should be nonzero;
(iii) in comparison to other goods, money should be relatively often
exchanged; (iv) the money lifetime should be sufficiently long
(lasting for many trades).

\begin{figure}[htb]
\centerline{%
\includegraphics[width=12.5cm]{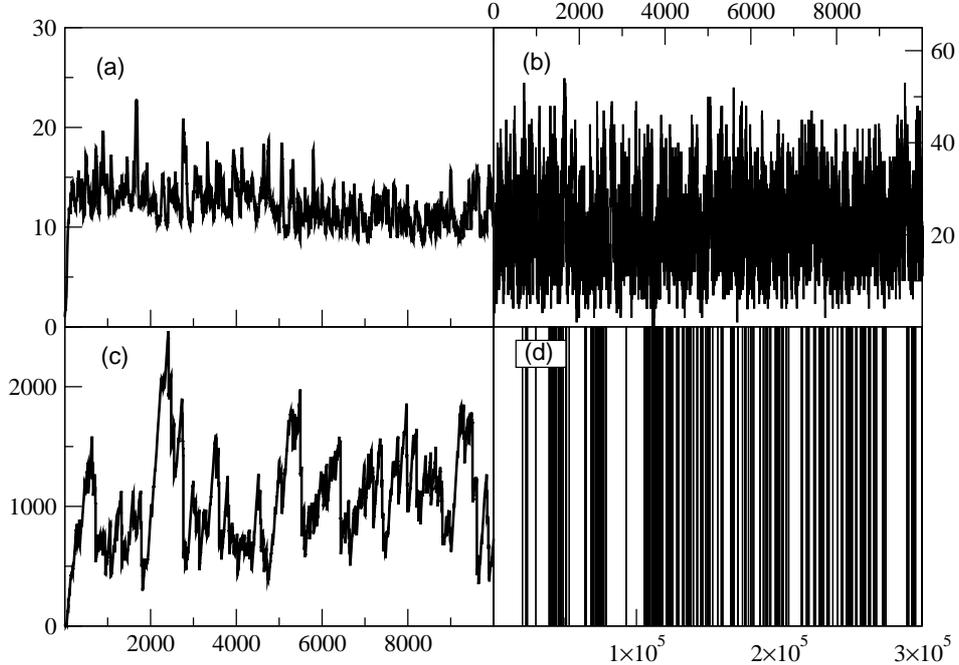}}
\caption{Model behavior for $N=50$ and threshold $Thresh = 2.5$.
(a) Time evolution of $V_{max}$;
(b) production of goods;  (c) supply of the most wanted good ("money");
(d) vertical lines indicate when the "money switching" takes place
for the first $10^4$ turns.}
\label{fig1}
\end{figure}

\begin{figure}[htb]
\centerline{%
\includegraphics[width=12.5cm]{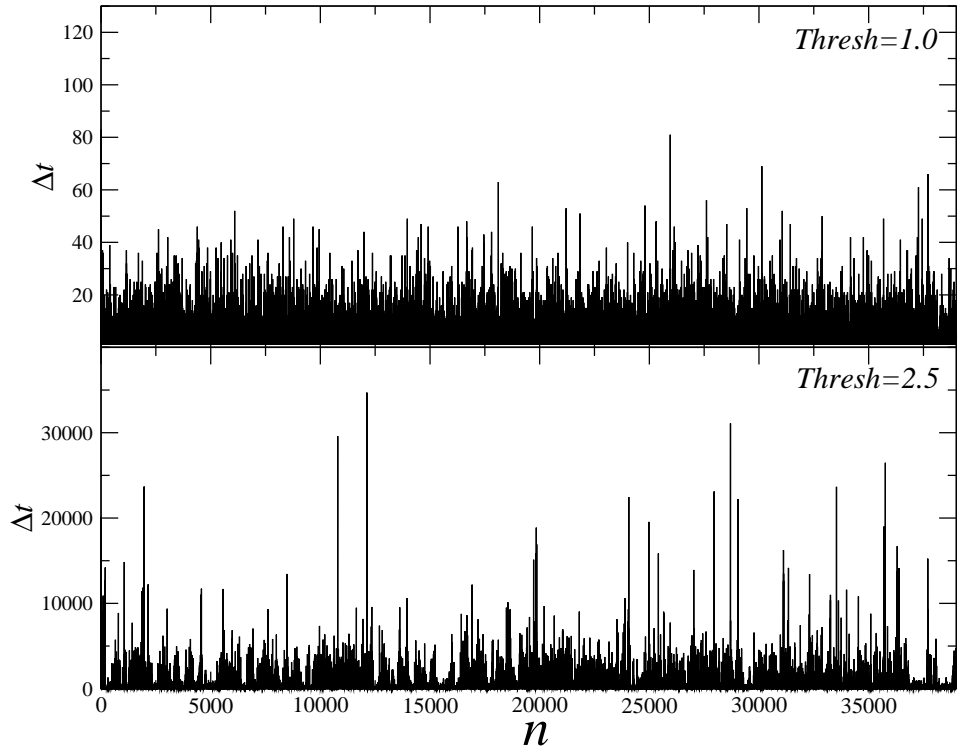}}
\caption{Consecutive "money lifetimes" measured in terms of the number of trading turns, during which
a given commodity fulfills criterion of being money, for two values of the model's threshold parameter $Thresh=1.0$ (top
panel) and $Thresh=2.5$ (bottom panel). Note different scales.}
\label{fig2}
\end{figure}

An example of time evolution of the principal characteristics of this model for the population size $N=50$ is shown
in Fig.~\ref{fig1} for the following dynamical variables: (a) $V_{max}$, (b) production of goods, (c) supply of the most
wanted good, and (d) points of the "money switching", i.e. the points when the most wanted good is overtaken by another
good.
The distances between the consecutive time-points are thus to be interpreted as "money lifetimes" within the present
model. These plots are given for the threshold value $Thresh=2.5$ which for $N=50$ marks a center of the region
where money start emerging. From the point of view of complexity science and in relation to the financial reality,
this is the most interesting region. The structure of "money lifetimes" seen in panel (d) of Fig.~\ref{fig1}
suggests their fractal organization and thus critical character of the transition from the phase with no money to the
money phase while increasing the parameter $Thresh$ value.

This is even more suggestive in Fig.~\ref{fig2} showing the consecutive "money lifetimes" for $Thresh=1.0$
and $Thresh=2.5$ measured in terms of the number of trading turns. While in the former case we see a white noise-like
structure with very short lifetimes, in the latter case the commodity fulfilling the criterion of money often happens to
reign for up to 35000 trading turns. The overall structure in this latter case displays a characteristic 'volatility
clustering', which is a hallmark of criticality and multiscaling \cite{lux1999,drozdz2009}. In the
following we therefore perform a more systematic quantitative study of this particular aspect of the present model using
the modern formalism of multifractality \cite{kwapien2012}.

\section{Multifractal analysis of "money switching" dynamics}

Multiscaling \cite{halsey1986,hentschel1983} represents a commonly accepted concept to grasp the most essential
characteristics of complexity. Indeed, the related measure in terms of multifractal spectra offers an attractively
compact frame to quantify the hierarchy of scales and specificity of their interwoven organization. This in particular
applies to the temporal aspects of complexity and, thus, well suits the present issue of "money switching" dynamics. Up
to now there exist two main types of algorithms to determine the multifractal spectra. The one that typically delivers
the most stable results \cite{oswiecimka2006} constitutes a natural extension of Detrended Fluctuation Analysis (DFA)
\cite{peng1994,kantelhardt2001} and is known as Multifractal Detrended Fluctuation Analysis (MFDFA)
\cite{kantelhardt2002}. The other algorithm - Wavelet Transform Modulus Maxima (WTMM)
\cite{muzy1994,arneodo1995} - is based on the wavelet decomposition of a signal. This algorithm requires
more care as far as stability of the result is concerned. It at the same time offers better visualization of the
relevant structures, however. For this reason, and also as a consistency test, below we  use both these
algorithms in parallel to analyze the same time series.

\subsection{MFDFA and results}

The MFDFA algorithm consists of several steps as sketched below. At first, for a time series $x_i$ a signal profile
is calculated according to the equation:
\begin{equation}
Y\left(j\right) =\sum_{i=1}^j[x_{i}-<x>]\ \  j=1...N ,
\end{equation}
where $< >$ denotes averaging over a time series of length $N$. Then the profile is divided into $2M_s$ disjoint
segments $\nu$ of length $s$ starting both from the beginning and from the end of the time series. For each box
the assumed trend is estimated by least-squares fitting a polynomial $P^{(m)}_{\nu}$ of order $m$.
Based on our own experience (O\'swi\c ecimka et al 2006) in the present analysis we use $m=2$ as optimal.
Next, variance of the detrended data is calculated:
\begin{equation}
F^{2}(\nu,s)=\frac{1}{s}\Sigma_{k=1}^{s}\lbrace Y((\nu-1)s+k)-P^{(m)}_{\nu}(k)\rbrace,
\end{equation}
and finally, by averaging the $F^{2}(\nu,s)$ function over all the segments $\nu$, a
$q$th-order fluctuation function is derived according to the equation:
\begin{equation}
F_q(s)=\lbrace\frac{1}{2M_{s}}\Sigma_{\nu=1}^{2M_{s}}[F^{2}(\nu,s)]^{q/2}\rbrace^{1/q},\ \
q\epsilon \Re \setminus \{0\}.
\end{equation}
In order to determine the statistical properties of the $F_q(s)$ function, the above
steps are repeated for different values of $s$. In the case of a fractal time series, fluctuation funtion $F_q(s)$
reveals power law dependence: $F_q\sim s^{h(q)}$, where $h(q)$ denotes the generalized Hurst exponent. For monofractal
time series, $h(q)$ is independent of $q$ and equals the well-known Hurst exponent $h(q)=H$. In the case of multifractal
correlations, however, $h(q)$ depends on $q$ and the Hurst exponent is obtained for $q=2$. From $h(q)$ exponents,
one can calculate the multifractal spectrum according to the equations: $\alpha = h(q)+qh^{'}(q) \quad \hbox{and} \quad
f(\alpha)=q[\alpha-h(q)]+1$, where $\alpha$ denotes the Hoelder exponent and $f(\alpha)$ is the fractal dimension of the
set of points with this particular $\alpha$. For multifractal time series, the singularity spectrum typically
assumes shape similar to an inverted parabola whose width is considered a measure of the degree of multifractality
and it thus shrinks to one point in the case of monofractal.

\begin{figure}[htb]
\centerline{%
\includegraphics[width=12.5cm,clip=true]{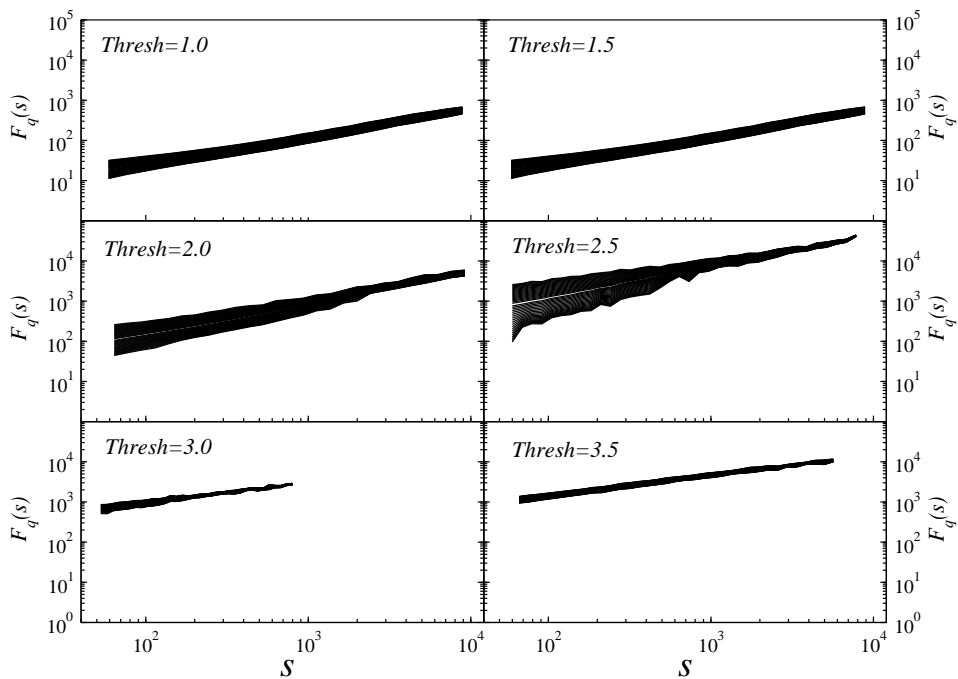}}
\caption{Fluctuation functions  $F_q(s)$ (for $q = -4\ldots +4$)
for the time series of money lifetimes generated from the model presented
for the threshold parameter ranging from $Thresh=1.0$ to $Thresh=3.5$
with the step of $0.5$.}
\label{fig3}
\end{figure}

By using this method, we then address the problem of correlations in the time series of money lifetimes. For this time
series (as the ones shown in Fig.~\ref{fig1}(d) or in Fig.~\ref{fig2}), the scale $s$ dependence for the set of the
so-evaluated fluctuation functions $F_q(s)$ ($-4 \le q \le +4$) around the critical threshold value for the
emergence of money, starting with $Thresh=1.0$ up to $Thresh=3.5$ with the step of 0.5, is shown in Fig.~\ref{fig3}. As
it can clearly be seen, these functions develop a power law form over the scale range of about two
orders of magnitude. However, the $q$-dependence of the corresponding power-law indices significantly varies with the
threshold parameter $Thresh$ and around $Thresh=2.5$ this dependence (here shown
for $-4 \le q \le +4$) is the most prominent, which signals nontrivial multifractality.
The degree of multifractality systematically increases while approaching this $Thresh$ value
from below, but then it sharply degrades when exceeding it.

\begin{figure}[htb]
\centerline{%
\includegraphics[width=12.5cm]{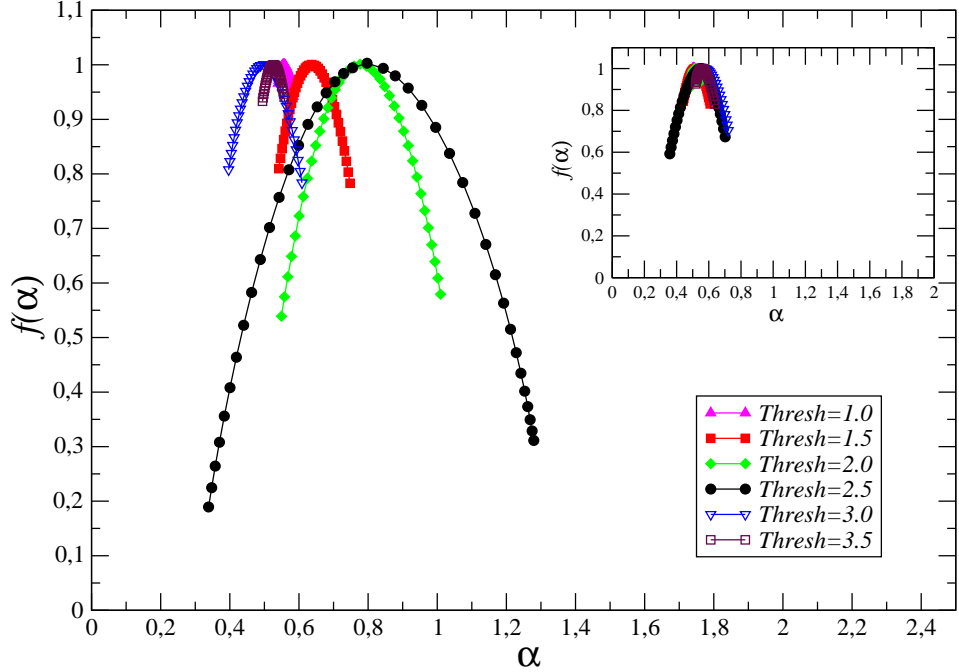}}
\caption{Singularity spectra $f(\alpha)$, calculated by means of MFDFA algorithm, for the same sequence of the original
time series of money lifetimes as in Fig.~\ref{fig3} and (inset) for the same but randomized (shuffled) time series of
money lifetimes.}
\label{fig4}
\end{figure}

The resulting singularity spectra $f(\alpha)$ are presented in Fig.~\ref{fig4} for the same sequence
of the parameter $Thresh$ values as in Fig.~\ref{fig3}. Indeed, the broadest spectrum - reflecting the hightest
complexity of the underlying dynamics \cite{kwapien2012} -  corresponds to $Thresh=2.5$
and departing from this value in either direction makes $f(\alpha)$ narrower with the maximum
moving towards $\alpha = 0.5$ like for the monofractal uncorrelated signals.
Comparison of these singularity spectra to the analogous quantities for the
randomized (shuffled) data, shown in the inset to Fig.~\ref{fig4}, clearly demonstrates significance of
this result. After such a destruction of temporal correlations, all the spectra become narrow and located in the
vicinity of $\alpha=0.5$. This comparison thus indicates that multifractal nature of the time series representing
consecutive money lifetimes is primarily due to the long-range temporal correlations.

\subsection{WTMM and results}

As already mentioned above, the WTMM method is an alternative technique of detecting the fractal properties of a signal.
The core of the algorithm is wavelet transform $T_{\psi}$ that is a convolution of a signal $x(i)$ and a wavelet $\psi$
\cite{grossmann1984}:
\begin{equation}
T_{\psi}(n,s)={1 \over s} \sum_{i=1}^{N}{\psi({i-n \over s}) x(i)},
\label{wavtrans}
\end{equation}
where $n$ denotes a shift of the wavelet kernel and $s$ is scale. In particular, the wavelet kernel $\psi$ can be
chosen arbitrarily. The only criterion is its good localization in space and in frequency domains. For the purpose of
our analysis, we choose the third derivative of the Gaussian $\psi^{(3)}(x)= {d^3 \over dx^3} (e^{-x^2/2}),$ because it
removes the trends that can be approximated by polynomials up to the second order. In the presence of singularity in the
data, the scaling relation of $T_{\psi}$ coefficients can be observed:
\begin{equation}
T_{\psi}(n_0,s)\sim s^{\ \alpha(n_0)}.
\end{equation}
However, this relation can be unstable in the case of densely packed singularities.
Therefore, it is suggested to identify the local maxima of $T_{\psi}$ and then to calculate the partition function
according to the equation:
\begin{equation}
Z(q,s)=\sum_{l \in L(s)} {|T_{\psi}(n_l(s),s)|^q},
\end{equation}
where $L(s)$ denotes a set of all maxima for scale $s$ and $n_l(s)$ is the position
of a particular maximum. In order to preserve the monotonicity of the family of the $Z(q,s)$ functions, the additional
condition needs to be imposed:
\begin{equation}
Z(q,s)=\sum_{l \in L(s)} {(\sup_{s'\le s} |T_{\psi}(n_l(s'),s')|)^q}.
\label{supremum}
\end{equation}
In the case of fractal signals, the $\tau(q)$ exponents characterize the power-law behavior of the partition function:
\begin{equation}
Z(q,s)\sim s^{\ \tau(q)}.
\end{equation}
For multifractal time series $\tau(q)$ is a nonlinear function of $q$, whereas it is
linear otherwise. The singularity spectrum is obtained by Legendre transforming $\tau(q)$ according to the
following formula:
\begin{equation}
\alpha=\tau'(q)\quad \textrm{and} \quad f(\alpha)=q\alpha-\tau(q).
\end{equation}

\begin{figure}[htb]
\centerline{%
\includegraphics[width=12.5cm]{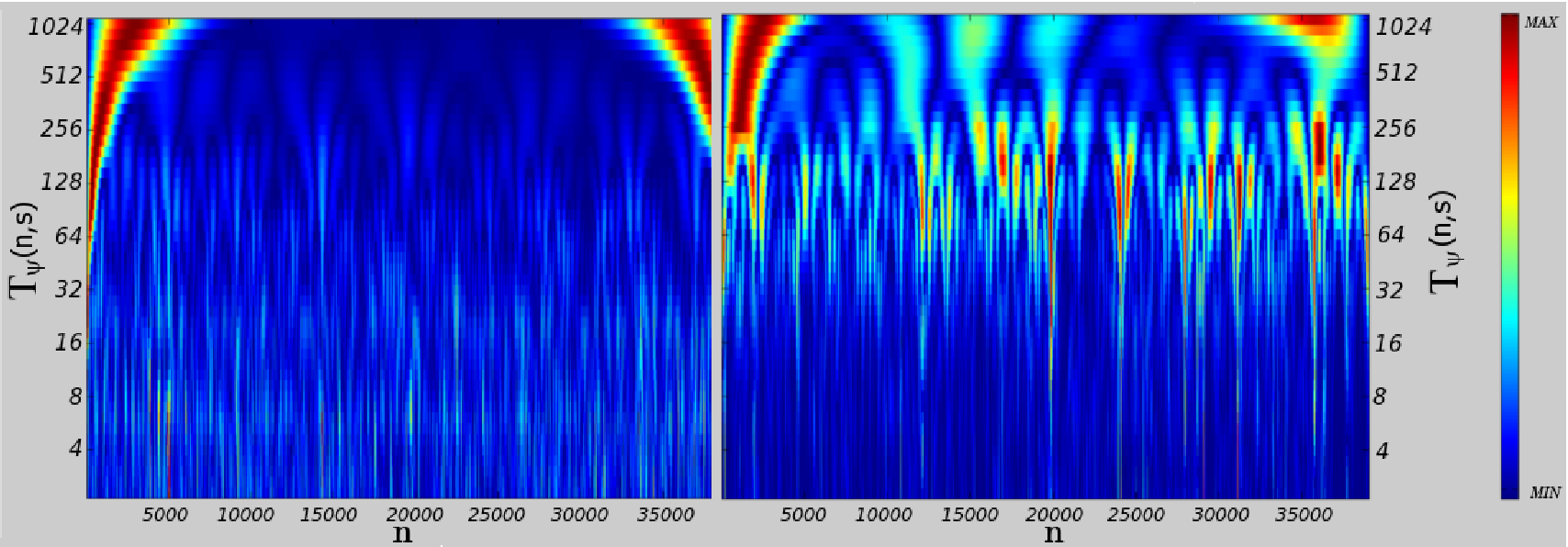}}
\caption{The wavelet transform $T_\psi$ of time series of money lifetimes. Left and
Right panel refers to time series generated with treshold value $Thresh=1.0$ and $Thresh=2.5$,
correspondingly. The wavelet coefficients were coded, independently at each scale, by
means of 128 colors ordered according to spectrum of natural light from blue (min
$T_\psi$) to red (max $T_\psi$). The wavelets used in calculation were third derivative
of Gaussian ($\psi^{(3)}$).}
\label{fig5}
\end{figure}

\begin{figure}[htb]
\centerline{%
\includegraphics[width=12.5cm]{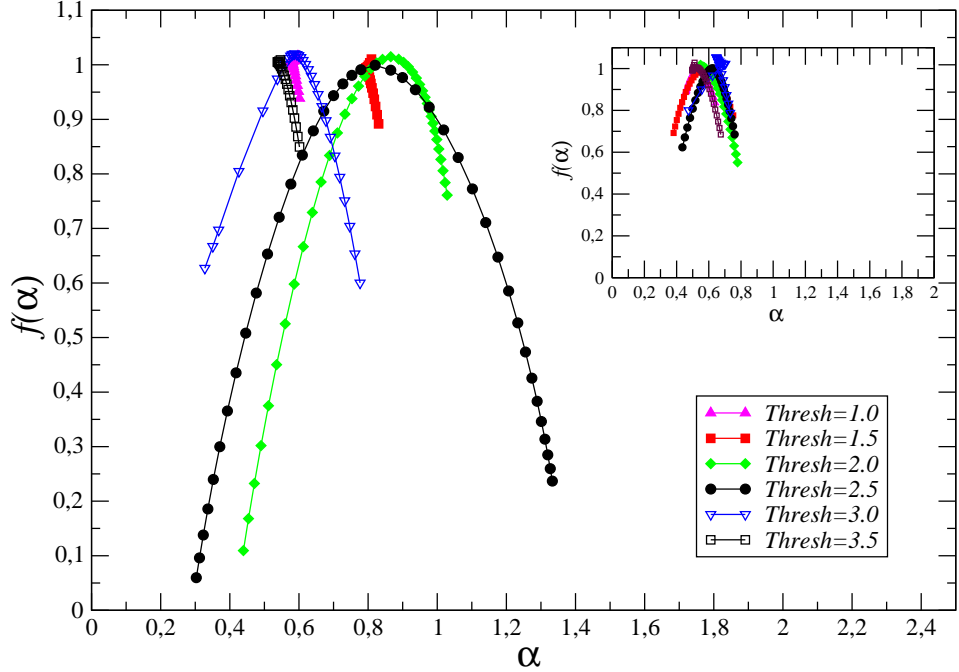}}
\caption{Singularity spectra $f(\alpha)$, calculated by means of WTMM algorithm, for the same sequence of the original
time series of money lifetimes as in Fig.~\ref{fig3} and (inset) for the same but randomized (shuffled) time series of
money lifetimes.}
\label{fig6}
\end{figure}

\begin{figure}[htb]
\centerline{%
\includegraphics[width=12.5cm]{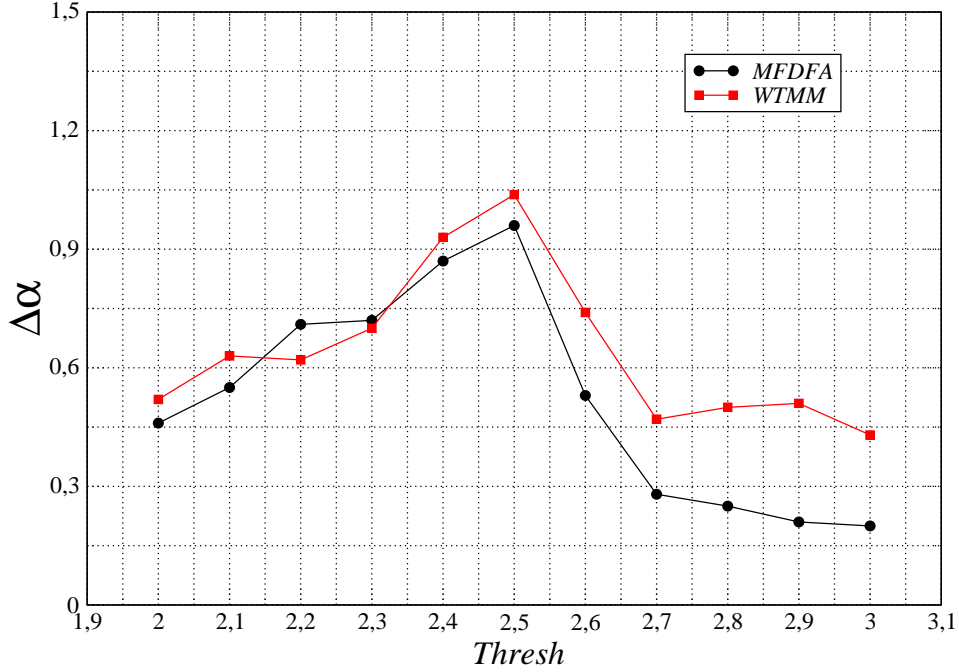}}
\caption{Width of singularity spectrum  $\Delta \alpha$ as function of model's threshold parameter. Black circles and
red squares refer to MFDFA and WTMM algorithm, respectively.}
\label{fig7}
\end{figure}

Two examples of the maps representing the wavelet transforms $T_{\psi}(n,s)$ calculated
from the same time series of money lifetimes as before for our model with $Thresh=1.0$ and $Thresh=2.5$ are shown
in Fig.~\ref{fig5}. The fractal character of these signals can be seen already on the visual level to manifest itself
quite convincingly from the bifurcation structure of the maxima of the wavelet transform.
For $Thresh=2.5$ (right hand side of Fig.~\ref{fig5}) the intensities of maxima vary with the scale, while for
$Thresh=1.0$ (left hand side of Fig.~\ref{fig5}) they remain largely homogenous, which signals a more involved fractal
composition in the former case. Of course, we already know from our previous MFDFA analysis that the $Thresh=2.5$ case
is strongly multifractal while the $Thresh=1.0$ case is essentially monofractal, and here we find an alternative
indication for this fact. Determining the singularity spectra according to the above-described WTMM algorithm for the
same six cases of the model's threshold parameter $Thresh$ as in Fig.~\ref{fig4} gives the results shown in
Fig.~\ref{fig6}. Comparing both MFDFA and WTMM results indicates basically the same tendency even on a fully
quantitative level: the broadest singularity spectrum characterizes the dynamics at around the critical threshold value
for the emergence of money. These comparisons are summarized globally in Fig.~\ref{fig7} in terms of the threshold
parameter dependence of the maximum span $\Delta \alpha$ of the corresponding singularity spectra obtained within the
MFDFA and WTMM methods. Both methods consistently point to $Thresh=2.5$ as this value of the model's threshold parameter
(for $N=50$) where the money switching dynamics is the most complex.

\section{Concluding remarks}

We have performed extensive numerical simulations using the agent-based computational economic model for the creation of
money, developed along the lines originally proposed by Menger \cite{menger1892} that money can spontaneously emerge in
a commodity exchange economy. Money in this model is defined as the most wanted good. A variant of this model studied in
the present
paper allows emergence as well as collapse of money. This model's ability is ruled by the only external parameter
$Thresh$ whose magnitude reflects agents tendency to act collectively and it induces memory effects. The most
interesting situation takes place just at the edge between a phase with no money emergence and a phase with stable
money. In this intermediate region one of the commodities spontaneously becomes universally wanted and retains such a
status for sufficiently long time so that it can be considered money. Under these edge conditions the dynamics of trade
is characterized by a permanent competition which leads to collapse of that particular money and another
commodity overtaking. An interesting related effect that we were witnessing at the course of simulations is that such
overtaking often alternates within one particular pair of commodities. In contemporary terms this can perhaps be
interpreted as this model's ability to induce effects in the spirit of competition of two currencies to become
world's leading (like USD versus Euro). From a theoretical point of view, the most interesting effect is that lifetimes
of the consecutive money emerging within the model at the edge, form time series that develop remarkable
multifractal characteristics as verified by the two independent algorithms: Multifractal Detrended Fluctuation
Analysis (MFDFA) and Wavelet Transform Modulus Maxima (WTMM). The width of the corresponding singularity spectra
(Fig.~\ref{fig7}) considered as a function of the threshold parameter resembles the $\lambda$-shaped
continuous phase transitions in physical systems, recently identified even in the complex networks dynamics~\cite{wilinski2015}.
That this transition can be paralleled to the critical phenomena is
primarily indicated by an increasing span of the multifractal spectrum (and thus an increasing complexity) when
approaching the transition point. The ability of the present model to generate the above effects seems very demanding
since the most creative acts of Nature are commonly considered to be associated with criticality \cite{bak1996}.
There also exists empirical evidence \cite{oswiecimka2005} that the consecutive intertransaction times on the stock
market form time series that are multifractal. One may speculate that there is some analogy between the present "money
lifetime" issue and the stock market intertransaction times: during a given "lifetime" money is the same;
similarly, price may change only during transaction. From this perspective intertransaction time can be considered
"price lifetime".

\end{document}